# Color-Doppler Echocardiography Flow Field Velocity Reconstruction Using a Streamfunction -Vorticity Formulation

**Authors:**

Brett A. Meyers[a], Craig J. Goergen[b], Patrick Segers[c], and Pavlos P. Vlachos[a,b*]

**Affiliations:**

[a]School of Mechanical Engineering, Purdue University, 585 Purdue Mall, West Lafayette, IN 47907, USA

[b]Weldon School of Biomedical Engineering, Purdue University, 206 S. Martin Jischke Dr., West Lafayette, IN 47907, USA

[c]bioMMeda Research Group, Institute Biomedical Technology (IBiTech), Ghent University, Ghent, Belgium

*Correspondence to: Pavlos Vlachos, School of Mechanical Engineering, Purdue University, pvlachos@purdue.edu

**Funding:**

This work was not supported by outside funding.

## ABSTRACT

We introduce a new method, (Doppler Velocity Reconstruction or DoVeR), for reconstructing two-component velocity fields from color Doppler scans. DoVeR employs the streamfunction-vorticity equation, which satisfies mass conservation while accurately approximating the flow rate of rotation. We validated DoVeR using artificial color Doppler images generated from computational fluid dynamics models of left ventricle (LV) flow. We compare DoVeR against the conventional intraventricular Vector Flow Mapping (iVFM$_{1D}$) and reformulated iVFM (iVFM$_{2D}$). LV model error analysis showed DoVeR is more robust to noise and probe placement, with noise RMS errors (*nRMSE*) between 3.81%-6.67%, while the iVFM methods delivered 4.16%-24.17% for iVFM$_{1D}$ and 4.06%-400.21% for iVFM$_{2D}$. We test the DoVeR and iVFM methods using *in-vivo* mouse-LV ultrasound scans. DoVeR yielded more hemodynamically accurate reconstructions, suggesting that it can provide a more reliable approach for robust quantification of cardiac flow.

# 1. INTRODUCTION

Resolving blood flow velocity vector fields in the left ventricle (LV) using ultrasound imaging has been a topic of research interest for over 30 years [1,2]. Two modalities are most often used: (1) B-mode ultrasound capturing flow tracer signals as grayscale speckle images, and (2) color Doppler ultrasound measuring blood velocity along transducer scan lines.

Echocardiogram particle image velocimetry (echoPIV) and blood speckle imaging (BSI) are block-matching methods used to track acoustic flow tracer signals in B-mode ultrasound images. EchoPIV uses acoustic-opaque bubbles as flow tracers, similar to conventional PIV tracers [3–5], and has been developed to improve reliability [6–9]. EchoPIV is not routinely performed, as the use of contrast agents for image enhancement is not clinically indicated unless initial image quality is poor. BSI uses plane wave imaging to generate speckle images of red blood cells [10] and because it is reliable at shallow depths [11], it is useful only for fetal and pediatric imaging.

Color Doppler imaging measures the blood velocity component along the ultrasound scan lines, producing two-dimensional (2D) maps of blood flow velocity values throughout the cardiac cycle. However, interpretation of flow patterns is a challenge as the scan gives an incomplete description of the underlying velocity vector field [12].

Echodynamography is one method developed for color Doppler vector field estimation, which separates the flow into 'base' and divergence-free flows to reconstruct the velocity vector field [13]. Although the 'base' flow is arbitrary and only performs well for rotating flows [14], it has been validated against PIV and used in research [15,16]. Pedrizzetti and Tonti introduced an alternative method using the irrotational flow assumption [17], but this formulation underestimates the strength of rotating flows [14]. Arigovindan *et al*. proposed a 2D reconstruction method based on registering Doppler views [18], which Gomez later extended to 3D [19].

The currently most used color Doppler velocity vector field reconstruction method for blood flow inside the LV is intraventricular vector flow mapping (iVFM). This method, which enforces the continuity equation [20], has the major drawback of oversimplifying the influence of wall and bulk fluid motion on locally reconstructed velocities [21], causing large, non-physical velocity gradients [22], requiring excessive smoothing. As a result, the method shows a significant error on the transverse velocity component, as high as 40% of the peak LV filling velocity [13,20]. Assi *et al.* introduced a generalized iVFM formulation [14], however, results still show elevated transverse component relative error (15-20%) during diastole. Lastly, both methods use free-slip or fixed-wall boundary conditions, but the former is not realistic, and the latter is invalid for large wall motion. Despite these issues, iVFM has been useful in research [23–28] and recently became commercially available by Hitachi [29]. Despite these advancements, adoption in clinical practice remains limited.

In this paper, we aim to improve upon the iVFM methods by providing a new color Doppler velocity vector field reconstruction methodology, which will deliver increased accuracy, robustness, and fidelity. Our novel algorithm is based on the kinematic equation relating the streamfunction ($\psi$) and vorticity ($\omega$) of the flow, which we term Doppler Velocity Reconstruction (DoVeR). In the following sections, we present the mathematical formulation of the method, perform validation and error analysis using artificial datasets, and demonstrate the tool in-vivo using mouse echocardiography scans. Results are compared against the conventional and generalized iVFM algorithms.

## 2. MATERIALS AND METHODS

### 2.1. Streamfunction-Vorticity Doppler Vector Reconstruction

Two-dimensional color Doppler imaging measures blood velocities, $v_D$, through a plane of interest. Each measurement is the projection of the real, three-component velocity vector $\vec{v} = \{v_x, v_y, v_z\}$ along the beam path, at pixel locations $x = (x, y, 0)$. We assume planar flow as the imaging is two-dimensional (2D). Hence, flow reconstruction must find $\vec{u} = [u_x, u_y, 0]$ that best estimates $\vec{v}$. Although this assumption introduces error [20,30], it still enables useful measurements [20,31]. We now introduce the streamfunction-vorticity Doppler vector reconstruction method.

The purely 2D streamfunction-vorticity (ψ-ω) formulation is well suited for axisymmetric flows such as those in blood vessels [32–34], cavo-pulmonary connections [35], and across cardiac valves [36,37]. The method is derived from the Navier-Stokes equations, and it is easy to implement and computationally effective [38]. In this work, we adapt the conservation of mass equation from this formulation.

Without loss of generality, we assign the measured velocity component $u_y = v_D$, such that, $u_x$ is the unknown velocity component that will be reconstructed. The measured velocity is preserved and remains unchanged throughout the algorithm. The vector $\vec{u}$ is reconstructed by computing the volume flux within the flow region using the streamfunction, $\psi$. The vector $\vec{u}$ is related to $\psi$ through the curl operator,

$$\vec{u} = \nabla \times \psi.$$    Equation 1

Since the flow is assumed 2D, $\boldsymbol{\psi}$ is a scalar, where $\boldsymbol{\psi} = [0,0,\psi]$. The rate of rotation, a measure of the angular velocity of the flow, is provided by the vorticity, $\boldsymbol{\omega}$, as,

$$\boldsymbol{\omega} = \boldsymbol{\nabla} \times \vec{\boldsymbol{u}}.$$

Equation 2

The planar flow assumption also allows $\boldsymbol{\omega}$ to be treated as a scalar, such that $\boldsymbol{\omega} = [0,0,\omega]$.

The vorticity and streamfunction are related by substitution of Equation 1 into Equation 2,

$$-\boldsymbol{\omega} = \boldsymbol{\nabla}^2 \boldsymbol{\psi}.$$

Equation 3

This is the ψ–ω equation, written in a scalar form as,

$$-\omega = \frac{\partial^2 \psi}{\partial x^2} + \frac{\partial^2 \psi}{\partial y^2}.$$

Equation 4

Equation 4 automatically enforces a divergence-free velocity field through the Laplace operator, while the vorticity is a function of the change in volume flux. Thus, the ψ–ω equation enforces the conservation of mass while relying on an angular velocity source term.

The DoVeR algorithm, which iteratively refines $\omega$, uses Equation 1, 2, and 4 to return a reconstructed velocity field. Figure 1 provides the algorithm flowchart. DoVeR begins by extracting velocities from color Doppler (CD), and pulsed-wave Doppler (PWD) scans. $u_y$ is set as the CD velocities. Doppler de-aliasing, when needed, is performed using a method of phase-unwrapping [39]. Each Doppler velocity is transformed into a phase angle within $\pm\pi$. The transformed velocity is unwrapped using guided flood filling based on phase quality, and if the unwrapped phase exceeds $\pm\frac{3}{2}\pi$ the initial phase is retained. The unwrapped field is converted back to units of velocity based on the velocity dynamic-range. An initial source term $\omega^0$ is quantified from $u_y$. The CD data can be used to automatically determine or manually draw boundaries, which aid in prescribing both the boundary conditions (BCs) along with PWD velocity measurements. The treatment of the BCs is given in the following section.

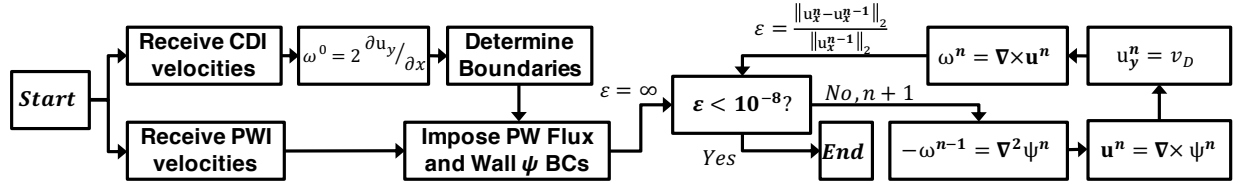

Figure 1: Flowchart of the Doppler velocity reconstruction (DoVeR) algorithm. The DoVeR algorithm requires color Doppler (CD) and pulsed-wave Doppler (PWD) scans as inputs to reconstruct underlying cardiovascular flow fields. CD data are used to determine initial source terms and boundaries. PWD data is used to determine boundary flux terms. These terms are coupled with the boundaries to construct appropriate BCs. The reconstruction solver is iteratively run to reconstruct *u*.

After setting $u_y$, $\omega^0$, and the BCs, an infinite error is provided to initialize the reconstruction process. At the beginning of each iteration, the error is checked against an error threshold, set at $10^{-8}$ for this work. This error threshold was chosen to deliver short computation times while ensuring the reconstructed velocity vector field doesn't change between iterations. The iteration error is computed from the L2-norm of the difference between the current and previous x-component velocities,

$$\varepsilon = \frac{\|u_x^n - u_x^{n-1}\|_2}{\|u_x^n\|}.$$

Equation 5

If the error is not below the threshold, the iteration number, *n*, is increased and a new value $\psi^n$ is computed from $\omega^{n-1}$ by LU-decomposition using the discrete formulation of Equation 4,

$$\ddot{\boldsymbol{D}}\psi^n = -\omega^{n-1},$$

<div align="right">Equation 6</div>

where $\ddot{\boldsymbol{D}}$ is the second-order derivative operator of size $N \times N$ with a 3-point stencil size, and $\omega^{n-1}$ and $\psi^n$ are $N \times 1$ vectors. Here $N$ is the total number of points within the reconstruction region. $u^n$ is calculated using the discrete formulation of Equation 1,

$$\mathbf{u^n} = [u_x^n \quad u_y^n] = [\dot{\boldsymbol{D}}_y\psi^n \quad \dot{\boldsymbol{D}}_x\psi^n],$$

<div align="right">Equation 7</div>

where $\dot{\boldsymbol{D}}_x$ and $\dot{\boldsymbol{D}}_y$ are the first-order derivative operators of size $N \times N$, with 3-point stencil size for derivatives in $x$ and $y$.

In order to further constrain the solver, we enforce that non-zero Doppler velocities $v_D$ must remain unchanged in $u_y^n$. We allow zero Doppler velocity replacement with discrete quantities from Equation 7. Zero Doppler velocity entries can be due to high pass filtering on ultrasound systems. We constrain replacement by thresholding to $\pm10\%$ of the maximum velocity.

Finally, $\omega^n$ is updated from $u^n$ using the discrete formulation of Equation 2,

$$\omega^n = \dot{\boldsymbol{D}}_x u_y^n - \dot{\boldsymbol{D}}_y u_x^n.$$

<div align="right">Equation 8</div>

and the procedure continues until convergence.

## 2.2. DoVeR Boundary Conditions

Equation 6 requires BCs for $\psi$ to obtain a physically consistent solution. LV flow has at least one inlet and/or one outlet, and one wall that need BCs. We use PWD velocity measurements at the mitral and aortic valve to set inlet and outlet BCs, ensuring the BCs are more physically consistent and making it possible to generalize the inlet and outlet flux BCs.

Otsu's threshold method is used to extract PWD velocity waveforms. The threshold image is evaluated to determine the maximum (or minimum) velocity at each time along the waveform to create a velocity time series. If more than one beat exists, we average all complete beat records to render a representative beat, which is then replicated to match the number of beats in the CD scan. We then subsample to match the CD frame rate.

We set $\psi_0 = 0$ along the wall separating the inlet and outlet, $\Omega_{sep}$, from $x_0$ to $x_b$ as shown in Figure 2, and we assume slug (uniform) flow for the inlet and outlet. For each frame, we assign the slug velocity profile based on the PWD mean velocity for $\mathrm{u}_{in}$ and $\mathrm{u}_{out}$. We integrate the velocity profiles outward from the separating wall to the wall of the LV, $\Omega_{wall}$, along a path, $s$,

$$\psi_{in}(t) = \int_{s_0}^{s_a} \mathrm{u}_{in}(t)\, ds + \psi_0, \qquad\qquad \text{Equation 9}$$

$$\psi_{out}(t) = \int_{s_b}^{s_c} \mathrm{u}_{out}(t)\, ds + \psi_0. \qquad\qquad \text{Equation 10}$$

Here the path $s$ is a function of Cartesian coordinates *x* and *y*. Simpson's rule is used for discrete integration. The difference between $\psi_{in}$ and $\psi_{out}$ is balanced as flux along $\Omega_{wall}$. We assume $\psi_{wall}$ as free penetration varying linearly as a function of position, $\psi_{in}$, and $\psi_{out}$,

$$\psi_{wall}(t) = f(x, y, \psi_{in}, \psi_{out}). \qquad\qquad \text{Equation 11}$$

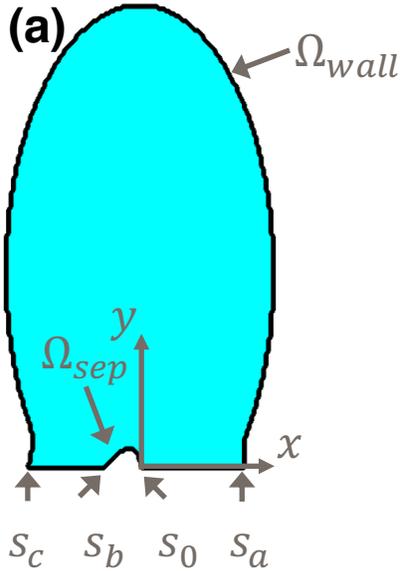

Figure 2: Diagram for imposing boundary conditions on (a) the artificial LV model.

## 2.3. Intraventricular Vector Flow Mapping

We implemented two versions of iVFM to benchmark against our DoVeR method. Both iVFM approaches are based around mass-conservation [20], resolving the azimuthal velocities from color Doppler images through the 2D polar coordinate continuity equation,

$$r \frac{\partial v_r}{\partial r} + v r + \frac{\partial v_\theta}{\partial \theta} = 0. \qquad \text{Equation 12}$$

The *conventional iVFM,* implemented as described in Garcia *et al.* [20], is referred to herein as iVFM$_{1D}$. The *reformulated iVFM,* a regularized least-squares formulation to reconstruct color Doppler velocity vector fields as described in Assi *et al.* [14], is referred to herein as iVFM$_{2D}$.

When employing either iVFM reconstruction methods, the Doppler scans are interpolated from a Cartesian coordinate system to a polar coordinate system. Implemented BCs for the iVFM methods are based on published literature.[14,20] For iVFM$_{1D}$, tangential flow (free-slip, impermeable) conditions were imposed along walls and normal flow along with inflow/outflow directions. For iVFM$_{2D,}$ the wall-normal velocity BCs was imposed.

## 2.4.  Doppler Ultrasound Artificial Data

The CFD velocity fields from a model of flow in the LV were provided by the bioMMeda research group at Ghent University [40,41]. We use the CFD velocity vector fields as the ground truth for the reconstruction method error analysis. The model heart rate is 60 beats per minute with 60% ejection fraction, volume varying between 170ml (end-diastole) and 69 ml (end-systole) [42].

Artificial color Doppler images were rendered from the CFD dataset using Field II [43,44]. Transducer settings were configured to match a phased array transducer, shown in Table 1, and simulated, as shown in Figure 3a. The probe was modeled with 128 elements, each generating a 2.5 MHz center frequency pulse with 4 pulse periods. The transmission focal point was set at a 65mm depth. A 7kHz pulse repetition frequency (PRF) was used to ensure no velocity aliasing. A total of 42,285 random scatterers were introduced (5 per cubic wavelength, $\lambda$=513$\mu$m) and ensonified to produce radiofrequency (RF) images. Doppler images (Figure 3b and 3c) were formed from RF signals using standard autocorrelation with a packet size of 10 per line [45]. PWD measurements for the mitral inflow and outflow track were sampled for inflow and outflow BCs.

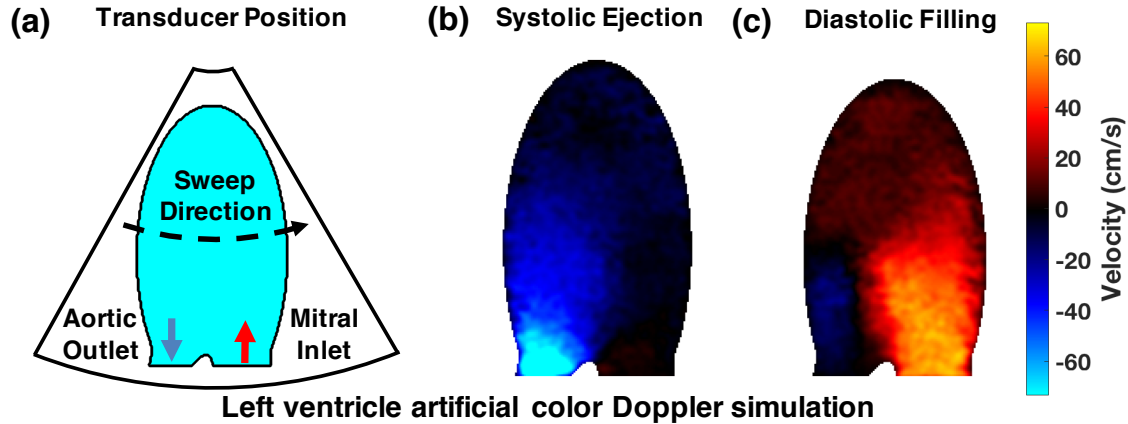

**(a)** Transducer Position  **(b)** Systolic Ejection  **(c)** Diastolic Filling

Aortic Outlet  Sweep Direction  Mitral Inlet

Velocity (cm/s)

**Left ventricle artificial color Doppler simulation**

Figure 3: Artificial color-Doppler ultras... . Example color-Doppler scans are shown for (b...

Tabl... ... settin...

| | |
|---|---|
| Num... | 128 |
| Image Line Density | 32/64/... |
| Transducer Center Frequency | 2.5 M... |
| Transmit Focal Point | 65 mr... |
| Transducer Element Height | 5 mm |
| Dynamic Receiver Focus | On |
| Transducer Excitation | Sinusoidal |
| Pulse Repetition Frequency | 7 kHz |
| Transmit Time ($\mu$s) | 1.6 |
| Packet Size | 10 |

Line density, probe orientation, added noise, and filtering strategy were varied across the artificial LV datasets to investigate their effects on velocity vector field reconstruction accuracy. Line densities of 32, 64, and 128 were used, mirroring low, medium, and high image resolution. Probe placement varied from 0° (apical long axis) to 90° (parasternal long axis) relative to the LV center. Gaussian white noise was applied across all test conditions; amplitude was adjusted from ideal to twice the expected level, or 0% to 40% of the model peak magnitude velocity [18,46,47]. Three filtering strategies were applied: no filtering, gaussian window filtering (GWF) with a 1.5mm standard deviation [20], and GWF with a 4.0mm standard deviation [22]. Velocity magnitude normalized root mean square error (nRMSE; by the CFD peak velocity magnitude) and velocity vector direction RMSE were quantified to determine reconstruction accuracy. Cumulative density functions (CDFs) of velocity magnitude normalized absolute error and vector direction absolute error was generated to observe error statistics.

## 2.5. Animal Ultrasound Data

The DoVeR and iVFM algorithms are demonstrated with *in vivo* imaging from small animal color Doppler ultrasound scans. All animal work was approved by the Purdue Animal Care and Use Committee (protocol # 1211000773). Imaging was performed on a Vevo2100 small animal ultrasound system (FUJIFILM VisualSonics Inc., Toronto, Ontario, Canada) with a 22-50MHz linear array probe (40 MHz center frequency; MS550D). One healthy male C57BL/6J wild-type mouse purchased from Jackson Laboratories (Bar Harbor, ME, USA) was examined at 21 weeks old. Apical long axis LV color Doppler scans for 60 distinct gated recordings were collected. Each recording was phase-locked at a time point in the cardiac cycle, providing a 50-frame image series across 50 different cycles. Each recording was averaged to generate a representative color flow image. PWD scans for the mitral inflow and outflow track were recorded. The LV was segmented by hand for each frame. The ultrasound settings are provided in Table 2. Although our demonstration is atypical of other Doppler reconstruction studies (shallow imaging depth, using traducers not typically used clinically, flow at Reynolds number one order of magnitude lower), flow through mouse and human hearts are characteristically similar.

Table 2: *in-vivo* Ultrasound Settings

| Color Doppler Settings |
| --- |

| | |
|---|---|
| Imaging Depth | 11 mm |
| Frame Size | 9 mm x 12 mm |
| Packet Size | 3 |
| Pulse Repetition Frequency | 20 kHz |
| **Pulsed-wave Doppler Settings** | |
| Range Gate Length | 0.5 mm |
| Range Gate Position | 8 mm (inflow) 8 mm (outflow) |
| Packet Size | 3 |
| Pulse Repetition Frequency | 32 kHz |

# 3. RESULTS

## 3.1. Artificial Data and Error Analysis

To investigate how probe placement and filtering affect accuracy, Figure 4 presents DoVeR, iVFM$_{1D}$, and iVFM$_{2D}$ velocity vector fields alongside the CFD ground truth during diastolic filling. Here we show the test case for 128-line density with 20% added noise. Instantaneous velocity vector fields are overlaid with the vorticity field, and vortex structures are shown as closed contours, using the coherent structure identification $\lambda_{CI}$ criterion with a 5% threshold of the swirl.[48]

In the standard apical long axis ($0^o$) orientation (Figure 4-i), flow is predominantly in the vertical direction along the scan lines. DoVeR (Figure 4-i-b) appears in good agreement with the ground truth (Figure 4-i-a); filtering shows no substantial improvement compared to the raw reconstruction. For iVFM$_{2D}$ (Figure 4-i-c), non-physical bands of vorticity and structures occur with no filtering (Figure 4-i-c-1). When the 4.0mm filter is applied, the iVFM$_{2D}$ accuracy improves. iVFM$_{1D}$ (Figure 4-i-d) shows agreement once the 4.0 mm filter has been applied, but without filtering, the reconstructions are dominated by noise.

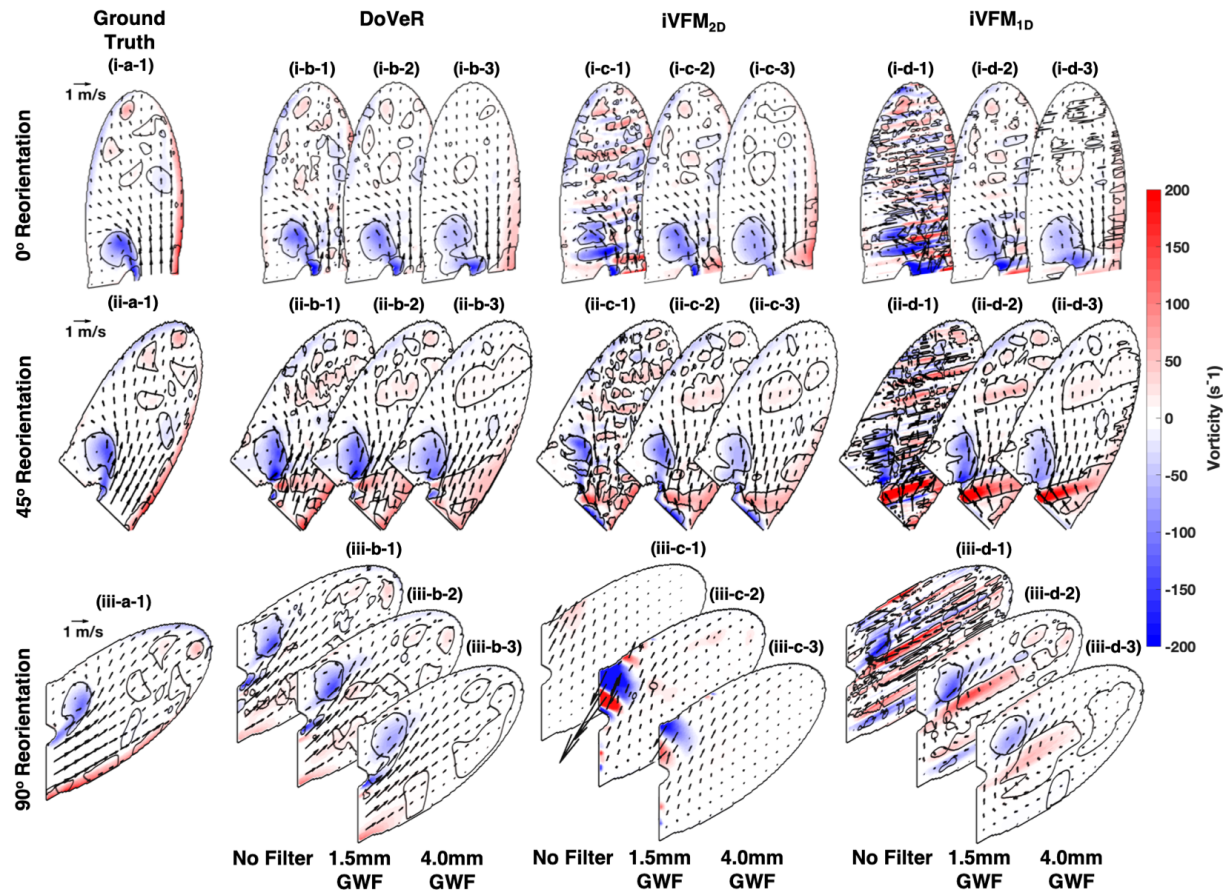

Figure 4: Early diastolic filling flow patterns for (a) the CFD model, (b) DoVeR, (c) iVFM$_{2D}$, and (d) iVFM$_{1D}$. Observations are presented for (i) 0º, (ii) 45º, and (iii) 90º probe reorientation angles; furthermore, for each reorientation, we present results when (1) no filtering, (2) 1.5mm Gaussian window filtering (GWF), and (3) 4.0mm GWF are applied. Closed contours represent vortex structures, identified using the $\lambda_{Ci}$ criterion, with a 5% threshold of the swirl.

The 45º orientation (Figure 4-ii) is more challenging, as the inflow is no longer perpendicular to the transducer face. DoVeR (Figure 4-ii-b) appears in good agreement with the ground truth (Figure 4-ii-a) and no improvement by filtering. iVFM$_{2D}$ (Figure 4-ii-c) under-resolves the inflow jet and renders a vortex that no longer retains its natural shape. iVFM$_{1D}$ (Figure 4-ii-d) shows no capability of resolving the inflow, and the flow is directionally dominant toward the transducer.

In the parasternal long-axis (90º) orientation (Figure 4-iii), flow is predominantly parallel to the transducer face. DoVeR (Figure 4-iii-b) appears in good agreement with the ground truth (Figure 4-iii-a). iVFM$_{2D}$ (Figure 4-iii-c) cannot resolve the flow, while iVFM$_{1D}$ (Figure 4-iii-d) cannot capture the inflow, an effect of the boundary conditions. iVFM$_{1D}$ still captures some of the vortex structures.

Figure 5 shows the RMS error and error CDFs on velocity magnitude and vector direction. Error on velocity magnitude was normalized by $\|\vec{v}_{CFD}\| = 1.53\ m/s$. All methods were mostly unaffected by resolution changes. DoVeR is presented as a line because results did not change (< 1% on velocity magnitude, < 5º on vector direction) with filtering.

Velocity magnitude *nRMSE* (Figure 5a) shows that DoVeR is as accurate as the smallest errors observed from iVFM$_{1D}$ and iVFM$_{2D}$ at 0º reorientation (*DoVeR:3.81%; iVFM$_{1D}$:4.16%, iVFM$_{2D}$:4.06%)*. The minimum error was achieved with the 1.5mm GWF; error increased 0.5% for iVFM$_{1D}$ and 2% for iVFM$_{2D}$ when the 4.0mm GWF was applied. When filtering is not applied, DoVeR outperforms iVFM$_{1D}$ and iVFM$_{2D}$, as noise increases. Under these conditions, iVFM$_{1D}$*, nRMSE* reaches 10.43%, and iVFM$_{2D}$ *nRMSE* reaches 13.65%. DoVeR error increases by ~3% as orientation angle changes from 0º to 90º. DoVeR minimum *nRMSE* is 3.81%, and maximum *nRMSE* is 6.67%. iVFM$_{1D}$ is affected by noise and orientation; however, the maximum error does not exceed 24.17%. iVFM$_{2D}$ is less affected by noise (*nRMSE*: 4.06% with 1.5mm GWF; 8.37% with no filtering); however, orientation significantly affects results. Once 30º orientation is achieved, error quickly grows and exceeds 400% at 90°. Smoothing improves results, but the error remains elevated compared to DoVeR (*nRMSE*: 13.65%).

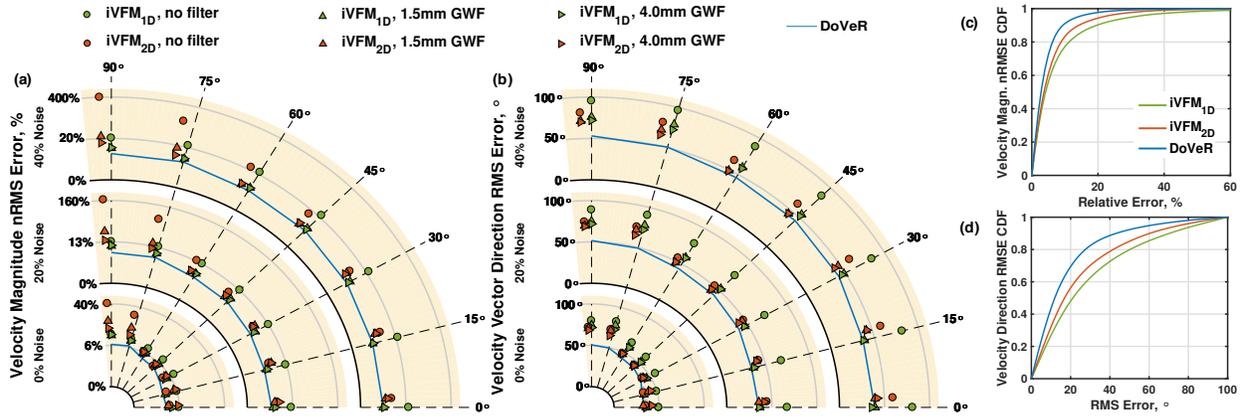

Figure 5: Artificial LV model RMS error for (a) velocity magnitude and (b) vector direction. Velocity magnitude is normalized by $\|\vec{v}_{CFD}\| = 1.54 \ m/s$. DoVeR, plotted as a solid line, is agnostic to image resolution changes, filtering effects, and noise. Quantitative error measurements are reported along the radial lines of the graph. Overall, DoVeR consistently provides more precise estimates compared to iVFM$_{2D}$ and iVFM$_{1D}$. Cumulative density function (CDFs) for (c) velocity magnitude and (d) vector direction support these findings.

Velocity vector direction RMSE (Figure 5b) shows that DoVeR is as accurate as the smallest errors observed from iVFM$_{1D}$ and iVFM$_{2D}$ at 0° reorientation (DoVeR:33.04°; iVFM$_{1D}$:35.84°, iVFM$_{2D}$:32.43°). Error increases by ~15° for DoVeR as orientation angle changes from 0° to 90° for each noise condition. DoVeR minimum *RMSE* is 33.80°, and the maximum *RMSE* is 53.15°. iVFM$_{2D}$ *RMSE* ranges from 32.43° to 81.06°, whereas iVFM$_{1D}$ *RMSE* ranges from 35.84° to 96.53°.

CDFs of velocity magnitude *nRMSE* (Figure 5c) and velocity vector direction *RMSE* (Figure 5d) provide reconstruction error probabilities across all conditions. DoVeR velocity magnitude errors are within 14% *nRMSE* and direction error within 61°. iVFM$_{2D}$ and

iVFM$_{1D}$ velocity magnitude errors fall within 22% and 31% *nRMSE*, respectively, and direction errors fall within 76° and 85°, respectively.

## 3.2.  In-vivo Data Analysis

Diastolic inflow for the mouse LV (Figure 6b) is expected to capture a dominant jet along the vertical direction with a pair of vortices along the jet exterior. The *in-vivo* scan (Figure 6b-1) captures the jet as the orange, high velocity (~0.3 m/s) region. The vortices are characterized by regions of dark blue, low velocity (< 0 m/s) near the jet.

DoVeR (Figure 6b-2), iVFM$_{2D}$ (Figure 6b-3), and iVFM$_{1D}$ (Figure 6b-4) reconstructions all capture the inflow jet. However, each method resolves the vorticity, vortex pair, and other flow structures with varying quality. DoVeR resolves one vortex in the pair as a complete structure, but the other vortex is split into two structures. The vorticity field captures a shear layer around the jet exterior and high vorticity near the vortex cores.

iVFM$_{2D}$ and iVFM$_{1D}$ are unable to resolve the vortex pair and instead capture multiple, non-continuous structures along with smaller structures. Both iVFM vorticity fields capture the shear layer around the jet exterior and high vorticity near the vortex cores. iVFM$_{1D}$ vorticity is still banded throughout the jet interior, which is not physically consistent.

Additional cardiac cycle time points are presented in Figure 6 at (a) isovolumic relaxation, (c) diastasis, and (d) systolic ejection. The systolic ejection phase (Figure 6d) is weak because the alignment of the scan is in the apical 2-chamber configuration, and the sampled frame is not well aligned with maximum outflow velocity.

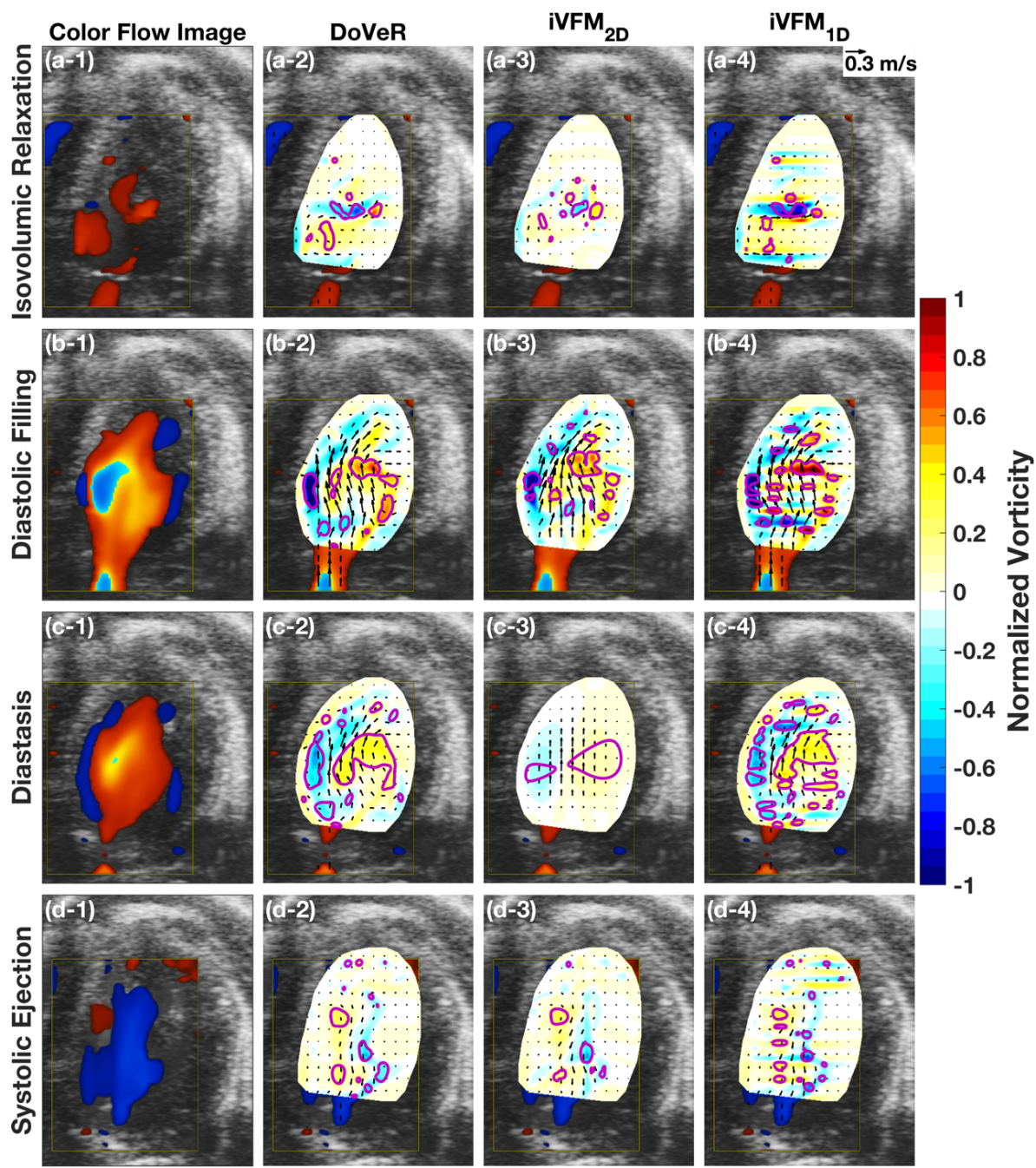

Figure 6: Demonstration of cardiac flow reconstruction of an *in vivo* color Doppler scan in a male wild-type mouse left ventricle. (1) Reference color Doppler scan used in flow reconstruction. Example reconstructions from (2) DoVeR, (3) iVFM$_{2D}$, and (4) iVFM$_{1D}$. Reconstruction presented at cycle times corresponding to (a) isovolumic relaxation, (b) diastolic filling, (c) diastasis, and (d) systolic ejection. Closed contours represent vortex structures, identified using the $\lambda_{ci}$ criterion, with a 5% threshold of the swirl.

## 4. Discussion

We introduce a new color Doppler velocity vector reconstruction algorithm, DoVeR, built on the kinematic equation relating the streamfunction and vorticity. Our formulation, which satisfies the conservation of mass, does not rely on primitive variables, unlike the iVFM algorithms. Error analysis with synthetic benchmark data and validation with in-vivo scans demonstrate that this modified formulation improves reconstruction accuracy. Primary differentiators contributing to error reduction include the numerical scheme for streamfunction-vorticity relationship, the simplified free-penetration BCs, and the iterative vorticity refinement scheme.

Our error analysis compared DoVeR and iVFM methods to study the accuracy of the numerical schemes and BCs, presented in Figure 4. DoVeR offers a nearly two-fold improvement in accuracy compared to the iVFM methods based on the error CDFs. As a result, DoVeR shows better performance across all test conditions, and the generalized BCs are more reliable compared to the iVFM BCs, reflected in the error CDFs.

iVFM$_{1D}$ integrates the continuity equation along a line to reconstruct transverse velocities, ensuring smooth estimates along the integration path. Velocity gradients between lines are still corrupted by noise, manifesting as "banded" fields, as seen in artificial LV vorticity (Figure 4) [13,22]. Two steps are employed to minimize noise, pre-processing the Doppler velocities by filtering and performing bi-directional integrations with weighted averaging.

iVFM$_{2D}$ relies on a least-squares method to solve the continuity equation. Four objective functions compose the cost function: a Doppler velocity constraint, the continuity equation, the BCs, and a smoothness constraint. Tikhonov regularization, which uses an L-curve corner search, determines the optimum weight of each term in the cost function. This generalized form of the iVFM$_{1D}$ offers smoothing, which should be unaffected by preprocessing filtering.

DoVeR relies on a Poisson PDE (Equation 4) which acts as a 2D area integral when solved numerically through LU-decomposition (Equation 6). The integration is inherently smooth, spreading the error to reduce noise, improving accuracy, and ensuring smooth velocity gradients, which should make filtering unnecessary.

Garcia *et al* [20] advise preprocessing Doppler velocities with a 3-pixel GWF, while Itatani *et al* [22] suggest the original GWF size is insufficient, advocating a 4.0 mm (8 pixels) as it does not significantly increase velocity error. Conversely, our study found the 4.0 mm GWF increases velocity magnitude and direction error. Both iVFM methods show (Figure 5) velocity magnitude error increases of 0.5% to 2%, and vector direction error increases $2^o$ to $10^o$ depending on test conditions. Additionally, the regularization used with iVFM$_{2D}$ fails when the input velocity is too smooth [49]. The reconstruction will under-resolve the velocity vector field, as observed in the animal imaging reconstructions (Figure 6). DoVeR does not show a considerable increase in velocity errors due to smoothing, although Figure 4 fields suggest that vortex identification is impacted.

Furthermore, only DoVeR truly satisfies the planar flow assumption. The Poisson PDE satisfies this assumption identically and enforces a divergence-free velocity field. The iVFM$_{1D}$ weighted averaging step assumes the flow is axisymmetric; however, path dependence of the line integral introduces errors due to the BCs and out-of-plane motion, which no longer preserves mass conservation. iVFM$_{2D}$ no longer enforces mass conservation once weighting is introduced.

The imposed BCs are critically important to the overall accuracy of the reconstructions. No-slip BCs enforce the heart wall and fluid, have the same velocity. Intracardiac reconstruction methods impose this by imposing B-mode speckle tracking velocities [19,21,29]. The wall tracking BC is viable in 3D reconstructions because wall motion is fully defined, but in 2D, this is not possible [14]. Garcia *et al*. adopt free-slip, tangential flow BCs, assuming the boundary layer is not resolvable [14,20], which is only practical for inertia-dominated flows. Assi *et al*. adopt normal flow BCs across the entire domain, which is appropriate when the inlet and outlet velocities are reliably measured.

Both iVFM methods are unable to reliably reconstruct flows when the dominant velocity component is no longer aligned in the axial direction. These methods expect the Doppler velocity to be aligned in the normal direction along permeable flow regions. When the dominant flow is close to or exactly parallel to the transducer face, this assumption is no longer valid. These results indicate that the iVFM methods are sensitive to probe misalignment and operator variability.

DoVeR uses Dirichlet conditions that are generalized for any domain where the inlet and outlet flux is conserved by flux along the walls. Inlet and outlet fluxes are defined by either PWD velocity measurements or by the color Doppler measurements as appropriate. This allows the BCs to be defined for any orientation. We assume the wall flux is uniform, although this is not always physically consistent. Despite this concern, DoVeR still shows improved velocity magnitude and vector direction accuracy and robustness to probe orientation.

Finally, vorticity plays an essential role in the DoVeR algorithm. In contrast to Pedrizzetti's method, which does not allow for vorticity production and underestimates the strength of rotating flows [14,17], DoVeR allows vorticity production by iteratively updating the unknown velocity component and constraining the known velocity component. As the vorticity iteratively updates, a smooth, divergence-free solution is achieved with a vorticity field that closely matches the underlying field, as demonstrated in Figure 4. Additionally, the refinement has a similar effect to the regularization coefficient determination used in iVFM$_{2D}$.

## 4.1. Limitations

The work presented here compares DoVeR and iVFM methods using artificial data, which mimics adult ultrasound and small animal ultrasound imaging of the LV. The animal imaging was performed at shallow depths using state-of-the-art probes not typically used in clinics. Human cardiac imaging differs as it is performed at greater depths and with phased array probes. Further investigation with human subject data is in progress and will be presented in future work.

DoVeR uses a piece-wise linear model for balancing mass flux along the walls as part of the LV BCs, which may over-simplify the wall motion. DoVeR's iterative vorticity refinement (1) replaces Doppler velocities that have been filtered due to machine high-pass filtering, and (2) allows vorticity production until the difference between iteration passes is minimized. Our current implementation prevents replaced Doppler velocities from exceeding 10% of the peak velocity when aliasing is not present to prevent vorticity over-production. This constraint is heuristic, therefore alternative strategies for replacement should be explored in future work.

## 5. CONCLUSION

We present DoVeR, a new algorithm for reconstructing the blood velocity vector in the LV from color Doppler ultrasound based on in-plane conservation of mass. DoVeR was compared against the conventional and reformulated iVFM methods using validation data of artificial color Doppler data rendered from a computational model of LV flow, Our analysis indicates DoVeR is robust to noise, probe placement, and filtering, providing RMS error on velocity magnitude and direction that never exceeds 6% and 53.15º, respectively. iVFM$_{1D}$ and iVFM$_{2D}$ normalized RMS error on velocity magnitude were 4.16%-24.17% and 4.06%-400.21%, respectively, while direction errors were 35.84º-96.53º and 32.43º-81.06º, respectively. Overall, DoVeR showed a nearly two-fold improvement in reconstruction accuracy compared to the iVFM methods. We demonstrated utility by reconstructing *in-vivo* data of LV flow from mouse scans. *In-vivo* results matched well with observations made within the validation data. DoVeR delivers more physically consistent reconstructions, providing a more reliable approach for robust quantification of cardiac flow.

## 6. ACKNOWLEDGMENT

Animal ultrasound data was collected by Arvin Soepriatna and Frederick Damen. The authors would like to thank Melissa Brindise for her assistance in the preparation and editing of this manuscript.

*Math.* **36**, 287–301.